\documentclass[12pt]{iopart}
\usepackage{graphicx,theorem}
\usepackage{setstack}
\usepackage{iopams}

\newcommand{\bu}{\mathbf{u}}
\newcommand{\bx}{\mathbf{x}}
\newcommand{\cM}{{\cal M}}
\newcommand{\cO}{{\cal O}}
\newcommand{\dv}{{\,\rm div \,}}
\newcommand{\orank}{{\,\mathrm{rank} \,}}
\newcommand{\Nor}{{\,\rm Nor \,}}
\newcommand{\bH}{\mathbf{H}}
\newcommand{\rot}{\mathrm{rot}\,}

\theoremstyle{plain}
\newtheorem{theorem}{Theorem}
\newtheorem{lemma}{Lemma}
\newtheorem{remark}{Remark}
\newtheorem{dfn}{Definition}

\eqnobysec

\begin{document}
\title[On the hierarchy of partially invariant submodels of differential equations]%
{On the hierarchy of partially invariant submodels of differential equations}

\author{Sergey V Golovin}

\address{Lavrentyev Institute of Hydrodynamics SB RAS,
Novosibirsk 630090, Russia}

\ead{sergey@hydro.nsc.ru}

\begin{abstract}
It is noticed, that partially invariant solution (PIS) of differential equations in many cases can be represented as an invariant reduction of some PIS of the higher rank. This introduce a hierarchic structure in the set of all PISs of a given system of differential equations. By using this structure one can significantly decrease an amount of calculations required in enumeration of all PISs for a given system of partially differential equations. An equivalence of the two-step and the direct ways of construction of PISs is proved. In this framework the complete classification of regular partially invariant solutions of ideal MHD equations is given.
\end{abstract}
 \ams{76M60, 58J70, 76W05, 35C05}


\submitto{\JPA}

\section*{Introduction}

The definition of a partially invariant solution (PIS) as a natural generalization of an invariant solution (IS) of differential equation was first suggested by L.V. Ovsiannikov \cite{LVO1,LVO}. At present there are known many examples of partially invariant solutions, mostly for models of fluid mechanics \cite{Nekitogi}--\cite{Thailert}. In contrast to the invariant solution, the partially invariant one is described by an overdetermined system of equations. The main difficulty of PIS construction is the investigation of compatibility and involutivity \cite{Pommaret} of this system. This complicated analysis requires individual approach for every particular solution. On the other hand, partially invariant solutions usually possess higher arbitrariness then the invariant ones, and, therefore, are more interesting for physical interpretation.

In the present work it is noticed that the majority of PISs of a given system of equations can be obtained as a result of the two-steps construction: As the invariant reduction of some partially invariant solution of the higher rank. This introduce a hierarchic structure in the set of all partially invariant solutions of a given system of equations. Those PISs which do not have a nontrivial two-steps representation are referred to as indecomposable ones. Analysis of all indecomposable PISs allows consequent construction of the whole set of partially invariant solutions by means of invariant reductions only. The latter do not require investigation of overdetermined systems of equation and, thus, are simpler. This investigation supplemented with LOT lemma \cite{LOT} introduces the hierarchic structure on the whole set of the group-invariant submodels of a given system of differential equations.

The theory of partially invariant solutions taken from \cite{LVO} is briefly recounted in sections 1 and 2. Hierarchy of partially invariant submodels is discussed in sections 3 and 4. Example of PISs hierarchy for shallow water equations is given in section 5. Section 6 is devoted to the complete classification of indecomposable regular PISs for ideal MHD equations.

\section{Partially invariant manifolds}
Let $G_r=\{T_a:\bar{x}=f(x,a)\}$, $a\in\Delta\subset\mathbb{R}^r$ be a local Lie group of transformations acting in space $x\in\mathbb{R}^n$. Basis of the corresponding Lie algebra $L_r$ of infinitesimal generators is given by
$X_\alpha=\xi_\alpha^i(x)\partial_{x^i}$, $(i=1,...,n$; $\alpha=1,...,r)$. Hereafter the Einstein summation convention on the repeating upper and lower indices is adopted. Let us observe a manifold $\cM$ regularly defined by equations
\begin{equation}\label{MPIS}
\cM:\;\;\psi^\sigma(x)=0,\;\;\;\sigma=1,\ldots,s;\;\;\;
\orank\bigl|\bigl|\partial\psi^\sigma/\partial x^k\bigr|\bigr|=s.
\end{equation}
Hereafter $\orank M(x)$ denotes the maximal rank of matrix $M(x)$ for various values of $x$.

\begin{dfn}
Orbit of point $x$ under $G_r$ group action is a set of points $\cO(x)=\{f(x,a)\;|\; a\in\Delta\subset\mathbb{R}^r\}$. Orbit ${\cal O}(\cM)$ of the manifold $\cM$ is the locus of orbits of all points $x\in\cM$: \[\cO(\cM)=\{f(x,a)\;|\;x\in\cM,\; a\in\Delta\subset\mathbb{R}^r\}\]
\end{dfn}
Let us introduce the following integer characteristic.
\begin{dfn}
Defect $\delta(\cM,G_r)$ of the manifold $\cM$ under $G_r$ group action is a difference between the dimensions of the orbit $\cO(\cM)$ and of the manifold $\cM$ itself:
\begin{equation}\label{defectPIS}
\delta(\cM,G_r)=\dim \cO(\cM)-\dim \cM.
\end{equation}
\end{dfn}
Defect of the manifold is an important characteristics showing the degree of non-invariancy of the manifold $\cM$ under the action of $G_r$. Defect of an invariant manifold is zero because orbit of each point of the manifold lies entirely on the manifold.

Orbit of an arbitrary manifold under the group action is itself an invariant manifold of the group because, by definition, it contains orbits of all its points. Moreover, orbit $\cO(\cM)$ is the minimal invariant manifold of the group $G_r$, containing $\cM$. Thus, it can be described in terms of the functional relations between the invariants of the group. Let the complete set of functionally independent invariants of $G_r$ be chosen in the form $I=\bigl(I^1(x),\ldots,I^t(x)\bigr)$, where $t=n-r_*$, and $r_*=\orank||\xi_\alpha^i(x)||$. By virtue of the theorem of representation of a non-singular invariant manifold \cite{LVO} the orbit of $\cM$ can be written in the form
\begin{equation}\label{PIS1}
\Phi^\tau\bigl(I^1(x),\ldots,I^t(x)\bigr)=0,\;\;\;\tau=1,\ldots,l.
\end{equation}

\begin{dfn}
Under the condition of regularity of specification (\ref{PIS1}), i.e. $\orank||\partial \Phi^\tau/\partial I^k||=l$ the number
\begin{equation}\label{rankPIS1}
\rho(\cM,G_r)=t-l
\end{equation}
is referred to as the rank of the partially invariant manifold $\cM$ with respect to the group $G_r$. Pair of integers $(\rho,\delta)$ define the type of the partially invariant manifold $\cM$.
\end{dfn}
Rank of the manifold is equal to the dimension of the orbit $\cO(\cM)$ in space of invariants of the group $G_r$. In practical calculations formula (\ref{rankPIS1}) is inconvenient, because it rely on the invariant representation (\ref{PIS1}) of the orbit of the manifold $\cM$, which may be not known explicitly. However, by using (\ref{defectPIS}), the rank can be found in terms of codimension $s$ of the initial manifold $\cM$ and of its defect $\delta(\cM,G_r)$:
\begin{equation}\label{rankPIS2}
\rho(\cM,G_r)=\delta(\cM,G_r)+t-s.
\end{equation}

\section{Partially invariant solutions}
Let us observe a system of differential equations
\begin{equation}\label{PDE1}
E:\;\;F^\sigma(x,u,\underset{1}{u},\ldots,\underset{k}{u})=0,
\;\;\;\sigma=1,\ldots,s.
\end{equation}
The main space is $Z=\mathbb{R}^n(x)\times\mathbb{R}^m(u)$. By $\underset{p}{u}$ we denote the set of all $p$-th order derivatives: $\left\{\frac{\partial^p u^j}{\partial x^{i_1}\ldots\partial x^{i_p}}\right\}$. System (\ref{PDE1}) admit Lie group $G_r=\{T_a:Z\times\mathbb{R}^r\rightarrow Z\}$. Action of $G_r$ can be prolonged on the derivatives in usual manner \cite{LVO, Olver}. Let Lie algebra of infinitesimal generators of $G_r$ be
\[ L_r=\{X_\alpha=\xi_\alpha^i(x,u)\partial_{x^i}+
   \eta_\alpha^k(x,u)\partial_{u^k},\; \alpha=1,\ldots,r\}. \]
For a $k$-dimensional subalgebra $H\subset L_r$ matrix
\[ H(\xi)=\left(
\begin{array}{ccc}
\xi_1^1(x,u)&\ldots&\xi_1^n(x,u)\\[2mm]
\ldots&\ldots&\ldots\\[2mm]
\xi_k^1(x,u)&\ldots&\xi_k^n(x,u)
\end{array}
\right), \] and extended matrix
\[H(\xi,\,\eta)= \left(
\begin{array}{cccccc}
\xi_1^1(x,u)&\ldots&\xi_1^n(x,u)&\eta_1^1(x,u)&\ldots&\eta_1^m(x,u)\\[2mm]
\ldots&\ldots&\ldots&\ldots&\ldots&\ldots\\[2mm]
\xi_k^1(x,u)&\ldots&\xi_k^n(x,u)&\eta_k^1(x,u)&\ldots&\eta_k^m(x,u)
\end{array}
\right).\]
are introduced. Let
\begin{equation}\label{SolRepr}
U:\;u^i=\varphi^i(x),\;\;\;i=1,\ldots,m.
\end{equation}
is a solution of equations (\ref{PDE1}).
\begin{dfn}
Solution $U$ is referred to as $H$-invariant solution ($H$-IS) of the system of equations $E$ if the manifold $U\subset Z$ is an invariant manifold under the subgroup $H\subset G_r$ action.
\end{dfn}
The necessary condition of $H$-IS existence is convenient to formulate in terms of the corresponding Lie subalgebra of infinetesimal generators.
\begin{lemma}\label{l1}
Lie subalgebra $H\subset L_r$ generate $H$-invariant solution of the system $E$ if the following equality holds
\begin{equation}\label{invarsolcond}
\orank H(\xi)=\orank H(\xi,\eta).
\end{equation}
\end{lemma}
Generalization of the notion of the invariant solution leads to the following.
\begin{dfn}
Solution $U$ is called $H$-partially invariant solution ($H$-PIS) of the system $E$ if the manifold $U\subset Z$ is a partially invariant manifold under $H\subset G_r$ action.
\end{dfn}
PISs are usually constructed on subalgebras of the admissible algebra which do not satisfy the necessary condition (\ref{invarsolcond}). Definitions of rank and defect of a partially invariant manifold are transferred naturally on PISs.  However, there is some specifics owing to the distinction of roles of $x$ and $u$ variables.

Let us introduce the following integer characteristics of $H$:

\bigskip
\begin{tabular}{lcp{88mm}}
$t=m+n-\orank H(\xi,\eta)$ &---&the total number of invariants of $H$;\\
$\sigma=n-\orank H(\xi)$ &---&number of invariants of group $H$, which depend only on $x$;\\
$\mu=t-\sigma$ &---&number of invariants essentially depending on $u$.
\end{tabular}

\bigskip
Partially invariant solution specified in form of a manifold $\Phi$ by formulae (\ref{SolRepr}) have codimension $s=m$. Rank of manifold $\Phi$ can be calculated by formula (\ref{rankPIS2}) as $\rho=\delta+t-m$. Orbit of manifold $\Phi$ is an invariant manifold of group $H$, therefore it may be specified by some functional relations of the form (\ref{PIS1}) for the invariants of the group. For the sake of simplicity we assume that the invariants are chosen in the separated form
\begin{equation}\label{PISInvars}
I:\left\{\begin{array}{l} I^1(x,u),\ldots,I^{\mu}(x,u),\\[2mm]
\lambda^1=I^{t-\sigma+1}(x),\ldots,\lambda^\sigma=I^t(x).
\end{array}\right.
\end{equation}
At that
\begin{equation}\label{rankcondinv}
\orank \frac{\partial \bigl(I^1(x,u),\ldots,I^\mu(x,u)\bigr)} {\partial
\bigl(u^1,\ldots,u^m\bigr)}=\mu.
\end{equation}

Let us form equations of the orbit of a partially invariant manifold $\Phi$ under the group $H$ action. At that it is required to specify the rank of a partially invariant manifold. This can be taken as any integer $\rho$, satisfying the inequality
\begin{equation}\label{rhoineq}
\sigma\le\rho<\min\{n,t\}.
\end{equation}
\begin{dfn}
Partially invariant solution is called regular if $\rho=\sigma$.
\end{dfn}
Equations of the orbit $\cO(\Phi,H)$ are constructed as the following set of functional relations between the invariants (\ref{PISInvars}) of the group $H$
\begin{equation}\label{PISInvpart}
\fl\Phi^\tau(\lambda^1,\ldots,\lambda^\sigma,I^1,\ldots,I^\mu)=0,
\;\;\;\tau=1,\ldots,t-\rho;\;\;\;
\orank\bigl|\bigl|\partial\Phi^\sigma/\partial I^k\bigr|\bigr|=t-\rho
\end{equation}
with unknown functions $\Phi^\tau$. In practical calculations this relations can be taken in the resolved form, e.g.
\begin{equation}\label{PISInvpartResolved}
I^\tau=\varphi^\tau(\lambda^1,\ldots,\lambda^\sigma,I^{t-\rho+1},\ldots,I^\mu),\;\;\;\tau=1,\ldots,t-\rho.
\end{equation}
Although, this form is more convenient for PIS computation, it is not unique in case of irregular solutions (i.e., $\rho>\sigma$). Therefore, in the theoretical analysis it is preferable to refer to the general form (\ref{PISInvpart}).

The dimension of the manifold (\ref{PISInvpart}) in the space of invariants is $\rho$. Equations (\ref{PISInvpart}) can be solved with respect to $t-\rho$ functions $u$ by virtue of conditions (\ref{rankcondinv}), (\ref{rhoineq}). Without loss of generality one can assume these functions to be  $u^1,\ldots,u^{t-\rho}$.  The remaining $\delta=m-t+\rho$ functions $u^{t-\rho+1},\ldots,u^m$ do not have representation in terms of invariants of $H$ and initially are not restricted by any extra assumptions. Thus, there appear $\delta$ non-invariant functions, which are assumed to depend arbitrarily on $x$.
\begin{equation}\label{NonInvRepr}
u^{t-\rho+1}=w^1(x),\ldots,u^m=w^\delta(x).
\end{equation}
Formulas (\ref{PISInvpart}), (\ref{NonInvRepr}) define the representation of the partially invariant solution of the type $(\rho,\delta)$.

\begin{remark}
In practical calculations it is necessary to observe all non-equivalent possibilities for solution of the orbit equations (\ref{PISInvpart}) with respect to functions $u$ and obtaining of the representation (\ref{NonInvRepr}) for non-invariant functions. In what follows the representation of solution will refer to the combination of the orbit equations  (\ref{PISInvpart}) with all possible representation of non-invariant functions of the form (\ref{NonInvRepr}).
\end{remark}

Substitution of the representation (\ref{PISInvpart}), (\ref{NonInvRepr}) of solution into the system of equations (\ref{PDE1}) leads to a factor-system of differential equations for the invariant functions $\Phi^k$, $k=1,\ldots,\mu$, and non-invariant functions $w^j$, $j=1,\ldots,\delta$. The factor-system of a partially invariant solution contains a subsystem $E/H$ for invariant functions and invariant variables, and equations $\Pi$ for the non-invariant functions. System $\Pi$ of equations should be observed as an overdetermined system for the non-invariant functions $w^j$. At that, all the invariant functions $\Phi^k$ are assumed to be known from a solution of the invariant subsystem $E/H$. The compatibility conditions of $\Pi$ usually extend both the invariant part $E/H$, and the system $\Pi$ itself. The purpose of investigation at this stage is to bring system $\Pi$ to involution, i.e. to obtain all its compatibility conditions to prove its self-consistency. Unfortunately, it is impossible to trace this process in general form. If this step is performed, the factor-system finally takes the form of a union of a subsystem $E/H$ for the invariant functions and of compatible on the solutions of $E/H$ system $\Pi$ for determination of non-invariant functions. This reduced system is simpler then the original system $E$ because $E/H$ involves less independent variables, and $\Pi$ contains less unknown functions.

\begin{dfn}
The union of the factor-system $E/H$ and of the system $\Pi$ will be referred to as $H$-partially invariant submodel of the system of differential equations $E$.
\end{dfn}

\section{Partially invariant solutions hierarchy} Let $H,N\subset G_r$ be subgroups, such that $H$ is a normal divisor in $N$: $H\lhd N$. Suppose that $H$ does not satisfy conditions (\ref{invarsolcond}) of invariant solution existence. Let $H$-partially invariant submodel of $E$ be known. In the sequel the following question is investigated: under which conditions on group $N$ there exists $N/H$-invariant solution for $H$-PIS and how does it relate to $N$-PIS of equations $E$?

Owing to the one-to-one correspondence between local Lie groups of transformations and their Lie algebras of infinitesimal generators \cite{LVO,Olver}, later on we do not distinguish between these two objects, denoting them by the same letter. All facts below proven in the Lie group language can be translated into the Lie algebras language and vice versa.

\begin{lemma}\label{l2}
The factor group  $N/H$ have induced action in the space of invariants of the group $H$.
\end{lemma}
{\sf Proof}. Factor group $N/H$ is a set of all left equivalence classes $g_l=g\circ H=\{g\circ h\;|\;h\in H\}$. Let $J$ be an invariant of the group $H$ action. Action of $g_l$ on $J$ is defined as $g_l (J)=g\circ h (J)=g(J)$, and, obviously, does not depend on the choice of the representative $h$ of the equivalence class. According to the condition $H\lhd N$ one has $g^{-1} H g\subset H$ for each $g\in N$. Let us show that if $J$ is invariant with respect to $H$ then with every $g\in N$ function $g(J)$ is also invariant under group $H$ action. Indeed, according to $g^{-1} \circ h\circ g (J)=h_1 (J)=J$ we have $h (g(J)) =g(J)$ for each $h\in H$. Thus, the factor group $N/H$ action is defined and closed in the space of invariants of group $H$. $\Box$

\begin{lemma}\label{l3}
Group $N/H$ is admitted by the system of differential equations $E/H$; group $N$ is admitted by the system $\Pi$ on solutions of $E/H$.
\end{lemma}
{\sf Proof}. Let us show that normal extension of group $H$ acts on the set of  $H$-invariant manifolds. Indeed, let $H\lhd N$ and $\cM$ is some $H$-invariant manifold. For any $h\in H$ and $g\in N$ there exists $h_1\in H$ such that $h\circ g=g\circ h_1$. Hence,
\[h(g(\cM))=g(h_1(\cM))=g(\cM).\]
Thus, the transformed manifold $g(\cM)$ is also $H$-invariant.

Orbit of $H$-PIS, given by equations (\ref{PISInvpart}), is an invariant manifold of the group $H$. Then, any transformation of group $N$ transforms an orbit of $H$-PIS into an orbit of $H$-PIS. The orbit of an arbitrary $H$-PIS is determined by the system $E/H$. Action of the factor-group $N/H$ is closed in the space of invariants of group $H$. Hence, transformations of factor-group $N/H$ acts on the set of solutions of $E/H$ system, i.e. are admitted by this system of differential equations. Next, group $N$ acts on the set of solutions of the original system $N$. Besides, it conserves an orbit of a solution in class of $H$-PIS. Thus, $N$ acts in the set of $H$-PISs, i.e. is admitted by system $\Pi$ on solutions of system $E/H$.  $\Box$

The necessary condition of $N/H$-invariant solution existence is convenient to formulate in the Lie algebraic language using infinitesimal generators of the observed Lie group.
\begin{lemma}\label{l4}
For $N/H$-invariant solution of the factor-system $E/H$ to exist the following condition should be satisfied
\begin{equation}\label{rankcond}
\orank N(\xi,\eta)-\orank N(\xi)=\orank H(\xi,\eta)-\orank H(\xi).
\end{equation}
\end{lemma}
{\sf Proof}. Let us transform infinitesimal generators of $H$ into the coordinate system, which flattens the group action:
\begin{equation}\label{newvars}
\begin{array}{ll}
y^1=\lambda^1(x),\ldots,y^\sigma=\lambda^\sigma(x);&y^{\sigma+1}=x^{\sigma+1},\ldots, y^n=x^n,\\[2mm]
v^1=I^1(x,u),\ldots,v^\mu=I^\mu(x,u);\;\;&v^{\mu+1}=u^{\mu+1},\ldots,v^m=u^m.
\end{array}
\end{equation}
Without loss of generality, this transformation is non-degenerate. Group $H$ acts transitively in the space of variables $(y^{\sigma+1},\ldots,y^n;\;v^{\mu+1},\ldots,v^m).$ Representatives of the factor-algebra $N/H$ have the form $\bar{X}=X+Y$, where $Y\in H$. Operator $Y$ does not contain any differentiations with respect to invariant variables $y^1,\ldots,y^\sigma$, $v^1,\ldots,v^\mu$. Operator $X$ by virtue of lemma  \ref{l2} projects into the space of invariants of group $H$, i.e. its coefficients at the differentiations with respect to invariant variables do not contain non-invariant variables. Thus, operator $Y$ does not participate in projection of operator $\bar{X}$ into the space of invariants of group $H$ at all. The Lie algebra of projections $\{X\}$ corresponds to the induced action of the factor-group $N/H$ in space of invariants of $H$. This Lie algebra should satisfy necessary conditions of the invariant solution existence (\ref{invarsolcond}). In what follows we check this conditions explicitly in terms of coordinates of infinitesimal operators of Lie group $N$.

Matrix of coordinates of infinitesimal generators of Lie algebra $N$ in $(y,v)$ coordinates have the block structure:
\[N(\xi,\eta)=
\left(\begin{array}{c|c||c|c}
0&A&0&B\\
0&0&0&C\\\hline\hline
K&L&R&S
\end{array}\right)
\]
First two columns of blocks (to the left of the vertical double line) correspond to differentiations with respect to independent variables $y$, the remaining two columns (to the right of the double line) are coordinates at differentiations with respect to dependent variables $v$. Number of columns in each block separated by vertical lines is equal to $\sigma$, $n-\sigma$, $\mu$, and $m-\mu$ correspondingly. The matrix horizontal division is such that first $\dim H$ rows (above the horizontal line) contain basic operators of Lie algebra $H$. The remaining operators (below the horizontal line) complete $H$ to $N$.

Blocks of matrix $N(\xi,\eta)$ with coordinates (row, column) $=(1,1)$, $(2,1)$, $(1,3)$, and $(2,3)$ are equal to zero because coordinates at differentiations with respect to invariant variables in infinitesimal generators of $H$ vanish. Algebra $H$ does not satisfy the necessary conditions of existence of an invariant solution (\ref{invarsolcond}). Hence, by performing a suitable combination of rows above the horizontal line in matrix $N(\xi,\eta)$, on can make the block with coordinates $(2,2)$ to be zero with $\orank A=n-\sigma$. As long as only the ranks of blocks are of interest, the linear combinations may be taken with coefficients depending on all variables. The resulting block $C$ is also non-degenerate: $\orank C=m-\mu$. Thus, blocks $A$ and $C$ are square non-degenerate matrices of dimensions $(n-\sigma)\times(n-\sigma)$ and $(m-\mu)\times(m-\mu)$.

Let us turn to the rows of $N(\xi,\eta)$ placed below the horizontal line and corresponding to basic elements of complement of $H$ to $N$. By a non-degenerate combination of these rows with upper $n-\sigma$ rows of matrix $N(\xi,\eta)$, one can zero block $L$. Next, by virtue of the non-degeneracy of block $C$, one can zero block $S$. At that, blocks $K$ and $R$ stay unchanged. Differential operators with coordinates from blocks $K$ and $R$ acting in the space of invariants of group $H$, are the required infinitesimal operators of induced action of the factor-group $N/H$ in the space of invariants. The necessary condition of existence of invariant $N/H$-solution (\ref{invarsolcond}) is that the rank of block $K$ is equal to the rank of matrix composed from blocks $K$ and $R$:
\begin{equation}\label{rankcondweak}
\orank K=\orank (K,R).
\end{equation}
This condition can be reformulated in terms of ranks of blocks of the complete matrix $N(\xi,\eta)$. Let columns of blocks of matrix $N(\xi,\eta)$ be denoted by Roman numbers $I$ -- $IV$. Notice, that by virtue of non-degeneracy of blocks $A$ and $C$ columns $II$ and $IV$ are linearly independent with each other and with columns $I$ and $III$.
\[
\begin{array}{l}
\orank(I,II)=\orank K+n-\sigma,\\[2mm]
\orank(I, II, III, IV)=\orank(K,R)+n-\sigma+m-\mu.
\end{array}
\]
Subtraction of the first equality from the second one gives
\[\orank(I, II, III, IV)-\orank(I,II)=\orank(K,R)-\orank K+m-\mu.\]
Relation (\ref{rankcondweak}) is satisfied if and only if
\[\orank(I, II, III, IV)-\orank(I,II)=m-\mu,\]
which is equivalent to (\ref{rankcond}) condition. As long as change of variables (\ref{newvars}) does not affect ranks of matrices, this condition can be verified in initial coordinate system. $\Box$

\begin{remark}
The necessary conditions formulated in lemmas \ref{l1} and \ref{l4} guarantee the possibility of construction of representations of the corresponding invariant solutions and finding factor-systems of equations for invariant functions. These conditions are not sufficient because compatibility of the obtained factor-systems can not be proved a priory.
\end{remark}

\section{Two-step construction of the solution}
The information obtained about partially invariant solutions allows formulating the following statement.
\begin{theorem}
Let system of differential equations $E$ admit Lie group of continuous transformations $N$. Suppose that there is a normal divisor $H$ in $N$, which does not satisfy the condition of existence of the invariant solution (\ref{rankcondinv}), but fulfils the requirement (\ref{rankcond}). Then, for the factor-system $E/H$, corresponding to $H$-PIS there exists an invariant solution with respect to the factor-group $N/H$. Moreover, the factor-system of the invariant solution $(E/H)/(N/H)$ is equivalent to the factor-system of the partially invariant solution $E/N$.
\end{theorem}

{\sf Proof}. The possibility of construction of $N/H$-invariant solution of $H$-PIS is already shown in lemmas \ref{l2}--\ref{l4}.
The only thing left to demonstrate is the equivalence of factor-systems $E/N$ obtained directly and by using the two-steps method as $(E/H)/(N/H)$.

Suppose, that the dimension of the factor-group $N/H$ is equal to $\kappa$. Owing to the condition (\ref{rankcondweak}) one can assume invariants $\lambda^1,\ldots,\lambda^\sigma$, and $I^1,\ldots,I^\mu$ of the group $H$ to be chosen in such a way that functions
\begin{equation}\label{InvVars}
\lambda^1,\ldots,\lambda^{\sigma-\kappa},\;\;I^1,\ldots,I^\mu
\end{equation}
form the basis of functionally independent invariants of the group $N$. This implies, that the action of the factor-group $N/H$ is transitive in the subspace $\mathbb{R}^\kappa(\lambda^{\sigma-\kappa+1},\ldots,\lambda^\sigma)$, whereas variables (\ref{InvVars}) are independent invariants of $N/H$. The representation of $N/H$-invariant solution of the factor-system $E/H$ is obtained by the demand of independency of functions $\Phi$ in (\ref{PISInvpart}) on variables $\lambda^{\sigma-\kappa+1},\ldots,\lambda^\sigma$. Exactly the same representation of the orbit of $N$-PIS of equations $E$ is obtained directly. Thus, representations of $N$-PIS and of $(E/H)/(N/H)$-IS coincide, hence the factor-systems followed are equivalent as well. $\Box$

Thus, in the set of partially invariant solutions of investigated system there is a hierarchic structure.
\begin{dfn}
Partially invariant submodel is called indecomposable if it can not be represented as the non-trivial combination of a partially invariant and invariant submodels.
\end{dfn}
Investigation of only indecomposable submodels allows significant reduction of efforts in enumeration of all partially invariant solutions of a given system of differential equations. Indeed, the most labor-intensive step of involutivity analysis for overdetermined systems of differential equations should be done only for indecomposable submodels. The remaining submodels are obtained from the indecomposable ones by means of only invariant reductions, which is usually much simpler. The following two sections demonstrate this approach on examples of the shallow water and ideal MHD equations.

\section{Shallow water equations} The equations, describing motions of a thin water layer over a flat bottom are observed:
\begin{equation}\label{shall2D}
E:\begin{array}{l}
u_t+uu_x+vu_y+h_x=0,\\[1mm]
v_t+uv_x+vv_y+h_y=0,\\[1mm]
h_t+(uh)_x+(vh)_y=0.
\end{array}
\end{equation}
Here $(u,v)$ is a particle's velocity vector, $h$ is the depth of the water layer. The basic space here is $\mathbb{R}^3(t,x,y)\times\mathbb{R}^3(u,v,h)$, hence $n=m=3$. The admissible algebra $L_9$ \cite{LVO1} is generated by operators (notations of paper \cite{Pavl} are adopted):
\[\begin{array}{l}
X_1=\partial_x,\quad X_2=\partial_y,\quad X_4=t \partial_x +\partial_u,\quad X_5=t\partial_y +\partial_v,\\[1mm]
X_9=x \partial_y-y\partial_x+u\partial_v-v\partial_u,\quad X_{10}=\partial_t, \\[1mm]
X_{11}=x\partial_x+y\partial_y+u\partial_u+v\partial_v+2h\partial_h,\\[1mm]
X_{12}=t^2\partial_t+tx\partial_x+ty\partial_y+(x-tu)\partial_u+(y-tv)\partial_v-2th\partial_h.\\[1mm]
X_{13}=2t\partial_t+x\partial_x+y\partial_y-u\partial_u-v\partial_v-2h\partial_h.
\end{array}\]

Let us observe a partially invariant solution given by Lie subalgebra \[N=\{X_1,\; X_4,\; X_{10}+X_{12}\}\subset L_9.\]
The subalgebra $H=\{X_1,\;X_4\}$ in $N$ is selected. It is easy to check that $H$ is ideal in $N$. Condition (\ref{rankcond}) is satisfied, therefore one can apply the two-steps algorithm.

The complete set of functionally independent invariants of $H$ is
\[t,\;y,\;v,\;h.\]
Here $n=3$, $m=3$, $t=4$, $\sigma=2$, $\mu=2$. Let us construct an indecomposable $H$-PIS of rank 2. Equation of orbit of a partially invariant solution (\ref{PISInvpart}) can be written in an explicit form
\[v=v(t,y),\;h=h(t,y).\]
Defect of the solution is $\delta=1$. There is only one non-invariant function $u$, which is supposed to depend on all independent variables: \[u=u(t,x,y).\]
Substitution of the obtained representation of solution into the initial system (\ref{shall2D}) gives the submodel equations. The first and the third equations of system (\ref{shall2D}) form overdetermined system $\Pi$ for the non-invariant function $u$.  From the third equation of (\ref{shall2D}) it follows that $u$ is linear on $x$:
\begin{equation}\label{shall2DNonInv}
u=k(t,y) x+U(t,y).
\end{equation}
At that, function $k$ have expression in terms of invariant functions: $k=-(h_t+vh_y)/h.$ For the sake of convenience, one can treat this expression as an additional equation of the factor-system $E/H$. Substitution of the representation (\ref{shall2DNonInv}) into the equations (\ref{shall2D}) and splitting with respect to $x$ leads to the system for invariant functions:
\begin{equation}\label{shall2DInv}
\begin{array}{l}
v_t+vv_y+h_y=0,\\[2mm]
h_t+vh_y+kh=0,\\[2mm]
k_t+vk_y+k^2=0
\end{array}
\end{equation}
and to equation for function $U$.
\begin{equation}\label{Pi}
U_t+vU_y+Uk=0.
\end{equation}
Equations (\ref{shall2DInv}) form the factor-system $E/H$, whereas equations (\ref{shall2DNonInv}), (\ref{Pi}) represent trivially consistent system $\Pi$ for the non-invariant function. The factor-system (\ref{shall2DInv}) itself admit some Lie group symmetries. According to lemma \ref{l3} the admissible group contains the subgroup with Lie algebra \[\frac{\Nor_{L_9}\{X_1,X_4\}}{\{X_1,X_4\}}=\{X_2,X_5,X_{10},X_{11},X_{12},X_{13}\}.\]
In particular, this algebra contains the subalgebra $N/H=\{X_{10}+X_{12}\}$. For construction of $N/H$-invariant solution of the $H$-PIS, operator $X_{10}+X_{12}$ should be re-written in terms of invariants of algebra $H$ in the following form (for convenience, it is also prolonged to the invariant variable $k$)
\[(t^2+1)\partial_t+ty\partial_y+(y-tv)\partial_v-2th\partial_h+(1-2tk)\partial_k.\]
Invariants of this operator are
\[\fl\lambda=y\sqrt{t^2+1},\quad V=v\sqrt{t^2+1}-t\lambda,\quad
   H=h(t^2+1),\quad K=k(t^2+1)-t.\]
Necessary condition of an invariant solution existence is obviously satisfied. Representation of the invariant solution of the factor-system (\ref{shall2DInv}) have the form:
\begin{equation}\label{VHKrepr}
v=\frac{V(\lambda)+t\lambda}{\sqrt{t^2+1}},\quad h=\frac{H(\lambda)}{t^2+1},\quad k=\frac{K(\lambda)+t}{t^2+1}.
\end{equation}
Substitution into the equation gives
\begin{equation}\label{shall2D_4}
\begin{array}{l}
VV'+H'=-\lambda,\\
VK'+K^2+1=0,\\
VH'+HV'=-KH.
\end{array}
\end{equation}
Equations (\ref{shall2D_4}) form a factor-system $(E/H)/(N/H)=E/N$ for the $N$-PIS of equations $E$. The corresponding system $\Pi$ is given by expression (\ref{shall2DNonInv}) and equation (\ref{Pi}) with substitution of invariant functions (\ref{VHKrepr}). For the integration of the obtained system new independent variable $\mu$ is introduced:
\begin{equation}\label{shall2D_5}
\frac{d\lambda}{d\mu}=V(\lambda),\quad \mu=\int\frac{d\lambda}{V(\lambda)}.
\end{equation}
Then, the second equation of (\ref{shall2D_4}) accurate to insufficient constant give
\begin{equation}\label{shall2D_6}
K=-{\rm tg} \mu.
\end{equation}
By using (\ref{shall2D_6}), one can integrate equation (\ref{Pi}):
\begin{equation}\label{Urepr}
 U=\frac{g(\mu-\arctan t)}{\cos\mu\sqrt{t^2+1}}
\end{equation}
($g$ is an arbitrary function). Besides, system (\ref{shall2D_4}) has a first integral, which follows from its third equation
\begin{equation}\label{7}
HV\cos\mu=m.
\end{equation}
Finally, system (\ref{shall2D_6}) possess a Bernoulli integral
\begin{equation}\label{Bern}
V^2+\lambda^2+4H=b_0.
\end{equation}
The latter should be observed as an implicit (not resolved with respect to the derivative) equation for the dependence $\lambda(\mu)$:
\begin{equation}\label{ImplEqn}
\bigl(\lambda'\bigr)^2+\lambda^2+\frac{4m}{\lambda'\cos\mu}=b_0.
\end{equation}
Hence, the $N$-PIS is finally given by expressions (\ref{shall2DNonInv}), (\ref{VHKrepr}), where functions $V$, $H$, $K$, and $U$ can be found form (\ref{shall2D_5})--(\ref{Bern}) after integration of the first-order ODE (\ref{ImplEqn}). Note, that the solution contains an arbitrary function $g$.

Analysis of the optimal system \cite{Pavl} for the 9-dimensional Lie algebra $L_9$ \cite{LVO1}, admitted by equations (\ref{shall2D}) shows that the combination of operators $\{\partial_x,\;t\partial_x+\partial_u\}$ or the equivalent combination $\{\partial_y,\;t\partial_y+\partial_v\}$ is presented in 9 three-dimensional representatives. All PISs of defect 1 generated by these subalgebras are decomposable and can be obtained by invariant reduction of equations (\ref{shall2DInv}) by means of one of the following operators:
\[\fl
\begin{array}{l}
X_2,\;\;X_{11},\;\;X_{10}+X_{11},\;\;X_5+X_{10},\;\;X_{10},\;\;aX_{11}+X_{13},\;\;X_5+X_{11}+X_{13},\\[2mm]
aX_{11}+X_{10}+X_{12},
\end{array}\]
Here $a$ is an arbitrary real parameter.

\section{MHD with general state equation} Equations of ideal magnetohydrodynamics \cite{KulikLubim,LandLifsh} are observed:
\[\begin{array}{l}
D\,\rho+\rho\dv\bu=0,\\
D\,\bu+\rho^{-1}\nabla p+\rho^{-1}\bH\times\rot\bH=0,\\
D\,p+A(p,\rho)\dv\bu=0,\\
D\,\bH+\bH\dv\bu-(\bH\cdot\nabla)\bu=0,\\
\dv\bH=0,\;\;\;D=\partial_t+\bu\cdot\nabla.
\end{array}\]
Here $\bu=(u,v,w)$ is a velocity vector, $\bH=(H,K,L)$ is the magnetic field; $p$ and $\rho$ are pressure and density. Thermodynamical functions are related by the state equation $p=F(S,\rho)$ with entropy $S$. Function $A(p,\rho)$ is determined by the state equation as $A=\rho\,(\partial F/\partial\rho)$. All functions depend on time $t$ and Cartesian coordinates $\bx=(x,y,z)$.

The admissible group is 11-dimensional Galilean group extended by homothety \cite{CRCII,Fuchs}. Infinitesimal operators form Lie algebra $L_{11}$ with basis
\[\fl\begin{array}{l}
X_1=\partial_x,\; X_2=\partial_y,\; X_3=\partial_z,\; X_4=t\partial_x+\partial_u,\; X_5=t \partial_y +\partial_v,\; X_6=t
  \partial_z +\partial_w,\\[1mm]
X_7=y \partial_z-z\partial_y+v\partial_w-w\partial_v+K\partial_L-L\partial_K,\; X_8=z\partial_x-x\partial_z+w\partial_u-u\partial_w+L\partial_H-H\partial_L,\\[1mm]
X_9=x\partial_y-y\partial_x+u \partial_v-v\partial_u+H\partial_K-K\partial_H,\;X_{10}=\partial_t,\;
X_{11}=t\partial_t+x\partial_x+y\partial_y+z\partial_z.
\end{array}\]
Optimal system of subalgebras $\Theta L_{11}$ was constructed in \cite{LVOSubm,GrundLala}; in the final form it can be found in \cite{LVOGDE03}. Regular partially invariant solutions for gas dynamics equations generated by representatives of this optimal system were investigated in papers \cite{Mel}--\cite{LVO&APCh2}. By virtue of special form of operators of Lie algebra $L_{11}$, partially invariant solutions of MHD are generated by the same representatives of the optimal system as in pure gas dynamics without the magnetic field. Hence, the known results can be taken into account in construction of the solutions for MHD equations. Only non-barochronous solutions will be observed below, i.e. such solutions that pressure $p$ depends on spatial variables $x$. Barochronous motions of fluid (pressure depend only on time) were carefully observed in general form in \cite{Chup1}--\cite{LVObaroh}.

As long as all operators of Lie algebra $L_{11}$ have nonzero coordinates at differentiations with respect to independent variables, one-dimensional subalgebras do not give rise to a PISs. The only two-dimensional subalgebra, which generate PIS of defect 1 and rank 3 is $\{X_1,X_4\}$.

Analysis of three-dimensional representatives of the optimal system shows that there are 15 subalgebras, responsible for PISs of defect 1 and rank 2. All of these solutions are generated by indecomposable PISs constructed on the following subalgebras
\begin{equation}\label{defect1rank2}
\begin{array}{l}\{X_2,\;X_3,\;X_7\},\;\;\;\{X_5,\;X_6,\;X_7\},\;\;\;\{X_7,\;X_8,\;X_9\},\\[1mm]
\{X_3+X_5,\;X_2-X_6,\;X_7\},\;\;\;\{X_3,\;X_5,\;X_2+X_6\}.
\end{array}
\end{equation}
All the rest of submodels of defect 1 and rank 2 can be obtained by the invariant reduction of $\{X_1,X_4\}$-PIS with respect to one of the following operators
\[\fl X_7+a X_{11},\;\;\;X_7+X_{10},\;\;\;aX_6+X_{11},\;\;\;X_5+X_{10},\;\;\;X_{10},\;\;\;X_2+X_6,\;\;\;X_6,\;\;\;X_2.\]

There are 31 four-dimensional representatives of the optimal system, which give rise to PISs of defect 1 and rank 1. These are representatives of $\Theta L_{11}$ with numbers 1, 4, 5 (at $\alpha=0$), 6, 7 (at $\alpha=0$), 9 (at $\beta=0$), 10 (at $\alpha=0$), 12--14, 16 (at $\alpha=0$), 17--21, 23, 29, 30, 35, 36, 38, 41--46, 48, 49 (numeration is given according to \cite{LVOGDE03}). There are only 9 indecomposable PISs among them with bases
\[\fl
\begin{array}{l}
\{X_1,X_5,X_6,\alpha X_4+X_7; \alpha\ne0\},\;\;\{\alpha X_1+X_4,X_5,X_6,\beta X_1+X_7;\beta\ne0\},\\[2mm]
\{X_1,X_2,X_3,\alpha X_4+X_7; \alpha\ne0\},\;\;\{\alpha X_1+X_4,X_3+X_5,X_2-X_6,\beta X_1+X_7;\beta\ne0\}\\[2mm]
\{X_2,X_3,X_4,X_1+X_7\},\;\;\;\{X_2,\alpha X_1+X_3,X_1+X_5,X_6;\alpha\ne0\},\\[2mm]
\{X_1,X_3+X_5,X_2-X_6,\alpha X_4+X_7;\alpha\ne0\},\;\;\;\{X_1,X_2,X_3+X_5,X_6\},\\[2mm]
\{X_1,\alpha X_2+\beta X_3+X_4,\;\sigma X_3+X_5,\tau X_2+X_6;\alpha^2+\beta^2=1,\alpha^2+\tau^2\ne0,
\beta^2+\sigma^2\ne0\}.
\end{array}
\]
All these subalgebras generate barochronous submodels, since their only invariant independent variable is time $t$. The only non-barochronous indecomposable partially invariant submodel is generated by a four-dimensional subalgebra $\{X_2,X_3,X_5,X_6\}$. The latter gives PIS of defect 2 and rank 2.

Among five-dimensional subalgebras of $L_{11}$ the only regular indecomposable and non-barochronous solution is generated by a subalgebra $\{X_2,X_3,X_5,X_6,X_7\}$. For MHD equations this solution have defect 3 and rank 2. All the rest of regular partially invariant solutions generated by higher-dimensional subalgebras of $L_{11}$ are either barochronous or decomposable.

The calculations above show that class of indecomposable regular non-barochronous PIS for MHD equations is exhausted by the submodels generated by a subalgebra $\{X_1,X_4\}$ (defect 1, rank 3), subalgebras (\ref{defect1rank2}) (defect 1, rank 2), subalgebra $\{X_2,X_3,X_5,X_6\}$ (defect 2, rank 2), and subalgebra $\{X_2,X_3,X_5,X_6,X_7\}$ (defect 3, rank 2). Note that partially invariant solutions generated by subalgebras $\{X_7,X_8,X_9\}$ and $\{X_2,X_3,X_7\}$ were already studied in \cite{GolovinSingVortMHD,arxiv0705,SingVortPlanar}.

\ack

The work was supported by Russian Foundation for Basic Research (project 05-01-00080), President programme of support of leading scientific schools and young scientists (grants Sc.Sch.-5245.2006.1, MK-1521.2007.1), and by Integration project 2.15 of Siberian Branch of RAS.

\end{document}